\documentclass[twocolumn, journal]{IEEEtran}

\IEEEoverridecommandlockouts

\usepackage{color}
\usepackage{graphicx}
\usepackage{amsmath}
\usepackage{amssymb}
\usepackage{algorithm}
\usepackage{algorithmic}
\usepackage{amsmath}
\usepackage{multirow}
\usepackage{booktabs}
\usepackage{array}
\usepackage{amsthm}
\usepackage{stfloats}
\usepackage{caption}
\usepackage{subfigure}
\usepackage{bm}
\usepackage{setspace}

\newcommand{\be}{\begin{equation}}
\newcommand{\ee}{\end{equation}}
\newcommand{\bea}{\begin{eqnarray}}
\newcommand{\eea}{\end{eqnarray}}
\newcommand{\ba}{\begin{array}}
\newcommand{\ea}{\end{array}}

\newcommand{\non}{\nonumber}

\allowdisplaybreaks[4]

\title{Joint Beamforming Design for RIS-Assisted Integrated Sensing and Communication Systems
\thanks{H. Luo, R. Liu, M. Li, and Y. Liu are with the School of Information and Communication Engineering, Dalian University of Technology, Dalian 116024, China (e-mail: luohonghao@mail.dlut.edu.cn; liurang@mail.dlut.edu.cn; mli@dlut.edu.cn; yangliu\_613@dlut.edu.cn).}
\thanks{Q. Liu is with the School of Computer Science and Technology, Dalian University of Technology, Dalian 116024, China (e-mail: qianliu@dlut.edu.cn).}}

\author{Honghao Luo,
        Rang Liu,~\IEEEmembership{Graduate Student Member,~IEEE,}
        Ming Li,~\IEEEmembership{Senior Member,~IEEE,}\\
        Yang Liu,~\IEEEmembership{Member,~IEEE,} and Qian Liu,~\IEEEmembership{Member,~IEEE}
        }

\begin{document}
\maketitle
\begin{abstract}
Integrated sensing and communication (ISAC) has been envisioned as a promising technology to tackle the spectrum congestion problem for future networks.
In this correspondence, we investigate to deploy a reconfigurable intelligent surface (RIS) in an ISAC system for achieving better performance.
In particular, a multi-antenna base station (BS) simultaneously serves multiple single-antenna users with the assistance of a RIS and detects potential targets.
The active beamforming of the BS and the passive beamforming of the RIS are jointly optimized to maximize the achievable sum-rate of the communication users while satisfying the constraint of beampattern similarity for radar sensing, the restriction of the RIS, and the transmit power budget.
An efficient alternating algorithm based on the fractional programming (FP), majorization-minimization (MM), and manifold optimization methods is developed to convert the resulting non-convex optimization problem into two solvable sub-problems and iteratively solve them.
Simulation studies illustrate the advancement of deploying RIS in ISAC systems and the effectiveness of the proposed algorithm.
\end{abstract}


\section{Introduction}
\thispagestyle{empty}
With the increasing demand for internet of everything (IoE) applications, spectrum has become a scarce resource, which stimulates the research on integrated sensing and communication (ISAC) to address the resulting spectrum congestion problem.
ISAC enables a fully-shared platform to transmit the dual-functional waveform for simultaneously performing the communication and radar sensing functionalities, which significantly improves the spectrum and hardware efficiencies.
As one of the crucial enabling technologies for future networks, ISAC has attracted widespread attention from both academia and industry \cite{F. Liu}-\cite{Tong JSTSP 2021}.

Plenty of researchers have engaged in designing the dual-functional waveforms for implementing ISAC.
The transmit beamforming design in ISAC systems is critical for achieving a better trade-off between the communication and radar sensing functionalities.
However, the performance improvement provided by the active beamforming design is very limited when facing severe channel degradations.
In order to tackle this problem, reconfigurable intelligent surface (RIS) has emerged as a revolutionary approach to provide additional degrees of freedom (DoFs) owing to its ability of intelligently tailoring propagation environments \cite{Wu TCOM 2021}-\cite{Wei TC 2021}.

RIS is generally a planar array consisting of numbers of passive, low-cost and hardware-efficient reconfigurable reflecting elements.
By cooperatively adjusting the reflecting coefficients, we can effectively create additional non-line-of-sight (NLoS) links to boost the system performance \cite{Wu TWC 2019}.
In light of these advantages, researchers have begun the studies of deploying RIS in ISAC systems \cite{Rang JSTSP 2021}-\cite{Jiang SJ 2022}.
The authors in \cite{Rang JSTSP 2021} maximized the radar signal-to-interference-plus-noise-ratio (SINR) under the communication quality-of-service (QoS) requirement and the power constraint.
The transmit waveform and the reflection coefficients are jointly designed to minimize the multi-user interference (MUI) for better communication QoS under the power constraint and the requirement of radar sensing in terms of beampattern similarity \cite{Wang TVT 2021a}.
However, it is unwise to simply minimize the difference between the received signals and the desired symbols, since the transmitter in ISAC systems usually has a relatively high power enabling much better communication QoS.

Motivated by the above findings, in this paper we investigate the joint active and passive beamforming design for the RIS-assisted ISAC system, where a multi-antenna base station (BS) simultaneously serves multiple single-antenna users with the aid of a RIS as well as detects multiple point-like targets.
We aim to maximize the achievable sum-rate under the beampattern similarity constraint for radar sensing, the restriction of the reflecting coefficients, and the total transmit power budget by jointly optimizing the transmit beamforming of the BS and the reflecting coefficients of the RIS.
In order to handle the resulting non-convex optimization problem, we first employ the fractional programming (FP) method to transform its objective function into a more tractable form, and then develop an efficient algorithm based on the the majorization-minimization (MM) and manifold methods to iteratively solve two subsequent subproblems.
Numerical results demonstrate the advantages of deploying RIS in ISAC systems and the effectiveness of the proposed algorithm.

\section{System Model and Problem Formulation}

We consider an ISAC system as depicted in Fig. 1, where a BS equipped with $M$ transmit/receive antennas simultaneously serves $K$ single-antenna users with the aid of an $N$-element RIS and detects $T$ point-like targets.
In particular, the RIS is deployed close to the users to effectively assist the downlink multiuser communications.
Meanwhile, since the RIS is deployed far away from the targets, which fly at low altitudes and have strong LoS links with the BS, the radar echo signals reflected via the RIS are relatively weak and negligible \cite{Wang TVT 2021a}.
\begin{figure}[htbp]
\centering
  \includegraphics[width=3.1 in]{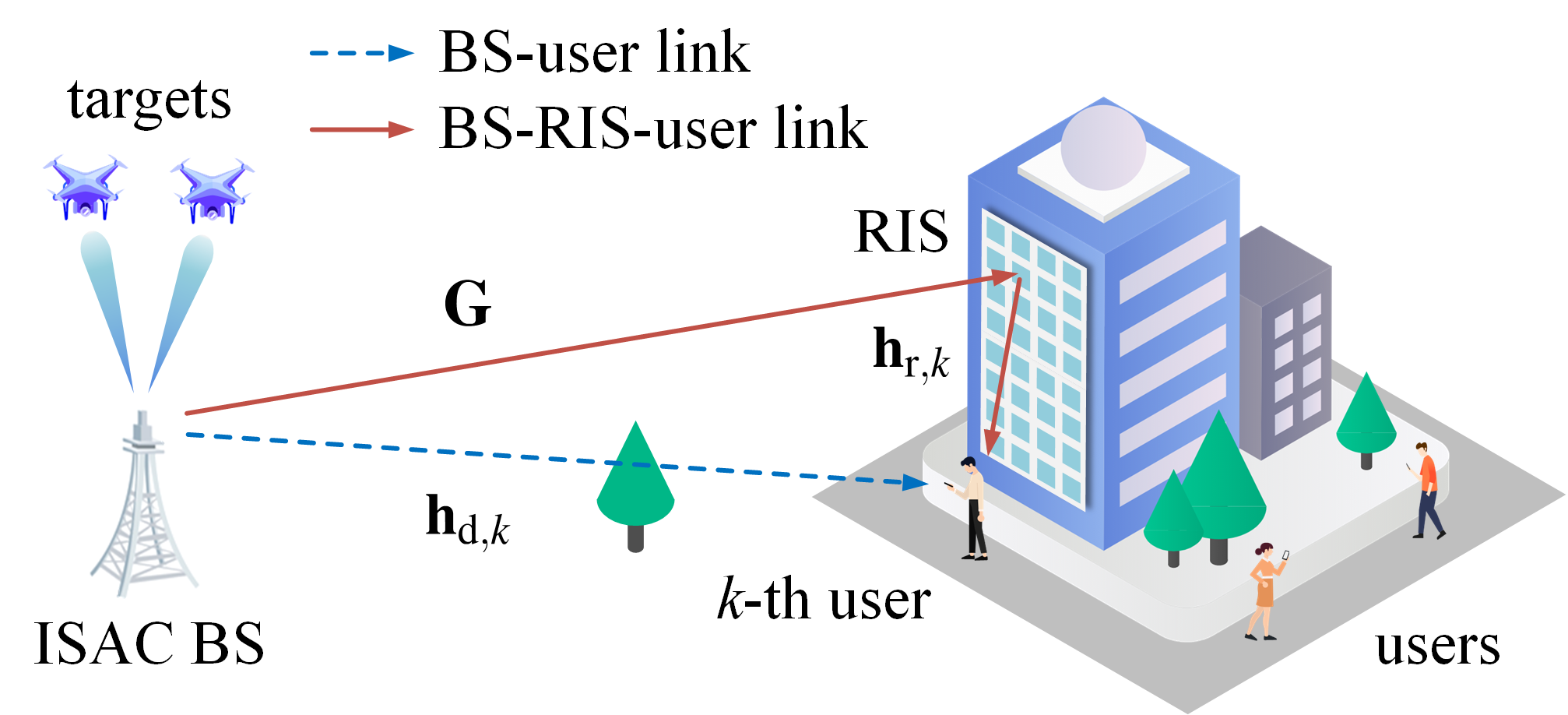}
  \vspace{0.2 cm}
  \caption{A RIS-assisted ISAC system.}
  \label{fig:system_model}
\end{figure}

The transmit signal of the BS is expressed as \cite{X. Liu}
\begin{equation}
\mathbf{x}=\mathbf{W}_\mathrm{c}\mathbf{s}_\mathrm{c}
+\mathbf{W}_\mathrm{r}\mathbf{s}_\mathrm{r}=\mathbf{W}\mathbf{s},
\end{equation}
where $\mathbf{W}_\mathrm{c}\in\mathbb{C}^{M \times K}$ and $\mathbf{W}_\mathrm{r}\in\mathbb{C}^{M \times M}$ respectively denote the communication beamforming matrix and the radar beamforming matrix.
Vector $\mathbf{s}_\mathrm{c} \in \mathbb{C}^{K}$ represents the communication symbol vector with $\mathbb{E}\{\mathbf{s}_\mathrm{c}\mathbf{s}_\mathrm{c}^H\}=\mathbf{I}_K$, vector $\mathbf{s}_\mathrm{r}  \in \mathbb{C}^{M}$ represents the radar probing signal with $\mathbb{E}\{\mathbf{s}_\mathrm{r}\mathbf{s}_\mathrm{r}^H\}= \mathbf{I}_M$, and they are assumed to be statistically independent of each other.
Specifically, we introduce the additional probing signal $\mathbf{W}_\mathrm{r}\mathbf{s}_\mathrm{r}$ to promote the sensing performance by exploiting more DoFs.
For brevity, we define the overall beamforming matrix as $\mathbf{W}\triangleq\left[\mathbf{W}_\mathrm{c}~ \mathbf{W}_\mathrm{r}\right] \in \mathbb{C}^{M \times (K+M)}$ and the transmit symbol vector as $\mathbf{s}\triangleq\left[\mathbf{s}_\mathrm{c}^T~ \mathbf{s}_\mathrm{r}^T\right]^T\in \mathbb{C}^{K+M}$.
Then, the received signal at the $k$-th user can be written as
\begin{equation}
y_k = ( \mathbf{h}_{\mathrm{r},k}^H\pmb{\Phi}\mathbf{G}+\mathbf{h}_{\mathrm{d},k}^H )\mathbf{x}+n_k,
\end{equation}
where $\mathbf{h}_{\mathrm{r},k} \in \mathbb{C}^{N}$, $\mathbf{G} \in \mathbb{C}^{N \times M}$ and $\mathbf{h}_{\mathrm{d},k} \in \mathbb{C}^{M}$ denote the baseband equivalent channels between the RIS and the $k$-th user, between the BS and the RIS, and between the BS and the $k$-th user, respectively.
The reflection matrix is defined as $\pmb{\Phi}\triangleq\mathrm{diag}(\pmb{\phi})$ where $\pmb{\phi}\triangleq[ \phi_1,\cdots,\phi_N ]^T$ with $|\phi_n|=1~,\forall n$.
The scalar $n_k\sim{\mathcal{C}\mathcal{N}(0,\sigma_k^2)}$ denotes the additive white Gaussian noise (AWGN) at the $k$-th user.
The SINR of the $k$-th user is thus given as
\begin{equation}
\mathbf{\gamma}_k = \frac{| ( \mathbf{h}_{\mathrm{r},k}^H\pmb{\Phi}\mathbf{G}+\mathbf{h}_{\mathrm{d},k}^H )\mathbf{w}_k |^2}{\sum_{j=1,j \neq k}^{K+M}| ( \mathbf{h}_{\mathrm{r},k}^H\pmb{\Phi}\mathbf{G}+\mathbf{h}_{\mathrm{d},k}^H )\mathbf{w}_j |^2+\sigma_k^2},
\end{equation}
where $\mathbf{w}_j$ is the $j$-th column of the beamforming matrix $\mathbf{W},j=1,\cdots,K+M$.
Then, the achievable sum-rate of all users is calculated as
\begin{align}
R=\sum_{k=1}^{K}\log_2(1+\gamma_k).
\end{align}

From the radar sensing perspective, in order to pursue better target detection and estimation performance, a widely adopted approach is to maximize the signal power in the directions of the targets and minimize it elsewhere.
In the sequel, the beampattern similarity metric that aims to match the designed beampattern with the ideal one is usually utilized to evaluate the sensing performance.
Specifically, we define the steering vector for direction $\theta$ as  $\mathbf{a}(\theta)\triangleq[ 1,e^{j 2\pi\frac{\delta}{\lambda} \mathrm{sin}(\theta)},\cdots,e^{j 2\pi\left( M-1 \right)\frac{\delta}{\lambda} \mathrm{sin}(\theta)} ]^T$, where $\delta$ is the antenna spacing and $\lambda$ denotes the signal wavelength.
Since the BS has complete knowledge of the communication symbols, $\mathbf{s}_c$ can also be utilized to boost the radar sensing performance \cite{X. Liu}.
The transmit beampattern, i.e., the signal power, can be expressed as \cite{J. Li}
\begin{equation}
\label{eq:transmit_beampattern}
P_\mathrm{b}(\theta;\mathbf{W})
=\mathbb{E}\big\{|\mathbf{a}^H(\theta)\mathbf{W}\mathbf{s}|^2\big\}
=\mathbf{a}^H(\theta)\mathbf{W}\mathbf{W}^H\mathbf{a}(\theta).
\end{equation}
The mean squared error (MSE) between the ideal beampattern and the designed beampattern, which evaluates the beampattern similarity, is thus given by
\begin{equation}
\mathcal{E}(\alpha,\mathbf{W})\triangleq\frac{1}{L}\sum_{l=1}^{L}\left|\alpha P_\mathrm{d}(\theta_l)-\mathbf{a}^H(\theta_l)\mathbf{W}\mathbf{W}^H\mathbf{a}(\theta_l) \right|^2,
\end{equation}
where normalized $P_\mathrm{d}(\theta_l)$ is the ideal beampattern, $\theta_l$ denotes the $l$-th sampled angle, and $\alpha$ is a scaling factor.
By introducing $\alpha$, the designed beampattern approximates the appropriately scaled ideal beampattern, instead of $P_\mathrm{d}(\theta_l)$ itself.

In this paper, we aim to jointly design the beamforming matrix $\mathbf{W}$ and the reflection coefficients $\pmb{\phi}$ to maximize the achievable sum-rate under the constraints of the beampattern similarity, the total transmit power, and the phase-shift of the RIS.
The optimization problem is thus formulated as
\begin{subequations}
\label{eq:initial_Problem}
\begin{align}
\max_{\mathbf{W},\pmb{\phi},\alpha}~~&\sum_{k=1}^{K}\log_2(1+\gamma_k)\\
\text{s.t.}~~&\frac{1}{L}\sum_{l=1}^{L} \left| \alpha P_\mathrm{d}(\theta_l)-\mathbf{a}^H(\theta_l)\mathbf{W}\mathbf{W}^H\mathbf{a}(\theta_l) \right|^2 \leq \epsilon,\\
&\| \mathbf{W} \|_F^2 \leq P,\\
&\left| \phi_n \right|=1,~\forall n,
\end{align}
\end{subequations}
where $\epsilon$ represents the level of the beampattern similarity and $P$ denotes the available transmit power.
It can be observed that problem (\ref{eq:initial_Problem}) is difficult to be optimized due to the objective function (\ref{eq:initial_Problem}a) with fractional and $\mathrm{log}(\cdot)$ terms, the quartic non-convex constraint (\ref{eq:initial_Problem}b), and the unit-modulus constraint (\ref{eq:initial_Problem}d).
Therefore, in the next section we first employ the FP  method to cope with (\ref{eq:initial_Problem}a).
Subsequently, convert problem (\ref{eq:initial_Problem}) into two tractable sub-problems and develop efficient algorithms to iteratively solve them.

\section{Joint Beamforming Design For RIS-Assisted ISAC Systems}

\subsection{FP-based Transformation}
In order to handle the complicated objective function (\ref{eq:initial_Problem}a), we first employ the FP method to convert it into a solvable polynomial expression.
Specifically, based on the Lagrangian dual reformulation \cite{K. Shen}, we introduce the auxiliary variable $c_k,~\forall k$, to take the ratio term $\gamma_k$ out of the $\log(\cdot)$ function and transform (\ref{eq:initial_Problem}a) into
\be
\label{eq:FP1}
\sum_{k=1}^{K}\log_2(1+c_k)-\sum_{k=1}^{K}c_k+\sum_{k=1}^{K}\frac{( 1+c_k )| \mathbf{h}_k^H\mathbf{w}_k |^2}{\sum_{j=1}^{K+M}| \mathbf{h}_k^H\mathbf{w}_j |^2+\sigma_k^2},
\ee
where for brevity we define the composite channel between the BS and the $k$-th user as $
\mathbf{h}_k^H\triangleq\mathbf{h}_{\mathrm{r},k}^H\pmb{\Phi}\mathbf{G}+
\mathbf{h}_{\mathrm{d},k}^H$.
The objective function (\ref{eq:initial_Problem}a) is equivalent to (\ref{eq:FP1}) when the auxiliary variable $c_k$ has the optimal value
\be
\label{eq:c_opt}
c_k^{\star}=\frac{| \mathbf{h}_k^H\mathbf{w}_k |^2}{\sum_{j=1,j \neq k}^{K+M}| \mathbf{h}_k^H\mathbf{w}_j |^2+\sigma_k^2},~\forall k.
\ee

Since the third multiple-ratio term in (\ref{eq:FP1}) still hinders a direct solution, we further apply the quadratic transform \cite{K. Shen} to convert it into
\be
\label{eq:FP2}
2\sqrt{1+c_k}\Re\{ g_k^{*}\mathbf{h}_k^H \mathbf{w}_k \}-| g_k|^2\sum\limits_{j=1}^{K+M}| \mathbf{h}_k^H\mathbf{w}_j |^2-| g_k|^2\sigma_k^2,
\ee
where the auxiliary variable $g_k$ has the optimal value
\be
\label{eq:g_opt}
g_k^{\star}=\frac{\sqrt{1+c_k} \mathbf{h}_k^H\mathbf{w}_k}{\sum_{j=1}^{K+M}| \mathbf{h}_k^H\mathbf{w}_j |^2+\sigma_k^2},~\forall k.
\ee

Based on the transformations in (\ref{eq:FP1}) and (\ref{eq:FP2}), after obtaining the auxiliary variables $c_k$ and $ g_k$, $\forall k$, we can recast the objective function (\ref{eq:initial_Problem}a) with respect to $\mathbf{W}$ and $\pmb{\phi}$ as
\be
\label{eq:after_FP}
\sum\limits_{k=1}^{K}\big(2\sqrt{1+c_k}\Re\{ g_k^{*}\mathbf{h}_k^H \mathbf{w}_k \}-\left| g_k \right|^2 \sum\limits_{j=1}^{K+M}\left| \mathbf{h}_k^H \mathbf{w}_j \right|^2\big).
\ee
Since the variables $\mathbf{W}$ and $\pmb{\phi}$ are coupled in the newly formulated objective function (\ref{eq:after_FP}), we propose to utilize the block coordinate descent (BCD) method to iteratively solve $\mathbf{W}$ and $\pmb{\phi}$ as presented in the following subsections.

\subsection{Optimize $\mathbf{W}$ with Given $\pmb{\phi}$}

With given $\pmb{\phi}$, the optimization problem of solving $\mathbf{W}$ can be formulated as
\begin{subequations}
\label{eq:optimize_W}
\begin{align}
\max_{\mathbf{W},\alpha}\,&\sum\limits_{k=1}^{K}\!\big(2\sqrt{1+c_k}\Re\{ g_k^{*}\mathbf{h}_k^H \mathbf{w}_k\!\}\!-\!| g_k |^2\!\!\! \sum\limits_{j=1}^{K+M}\!\!\!| \mathbf{h}_k^H \mathbf{w}_j |^2\big)\\
\text{s.t.}~\,&\frac{1}{L}\sum_{l=1}^{L} \left| \alpha P_\mathrm{d}(\theta_l)-\mathbf{a}^H(\theta_l)\mathbf{W}\mathbf{W}^H\mathbf{a}(\theta_l) \right|^2 \leq \epsilon,\\
&\| \mathbf{W} \|_F^2 \leq P.
\end{align}
\end{subequations}
It can be easily found that the variable $\alpha$ only exists in constraint (\ref{eq:optimize_W}b), of which left-hand side is a quadratic and convex function with respect to the variable $\alpha$.
Therefore, we can easily obtain its minimum by using the typical first-order optimality condition, i.e., $\frac{\partial\mathcal{E}(\alpha,\mathbf{W})}{\partial\alpha}=0$.
The optimal $\alpha$ is then calculated as
\be
\label{eq:alpha_opt}
\alpha^{\star}=\frac{\sum_{l=1}^{L}P_\mathrm{d}(\theta_l)
\mathrm{vec}^H(\mathbf{A}_l)\mathrm{vec}(\mathbf{W}\mathbf{W}^H)}
{\sum_{l=1}^{L}P_\mathrm{d}^2(\theta_l)},
\ee
where we define $\mathbf{A}_l \triangleq \mathbf{a}(\theta_l)\mathbf{a}^H(\theta_l)$ for simplicity.
Substituting the optimal $\alpha^{\star}$ (\ref{eq:alpha_opt}) into $\mathcal{E}(\alpha,\mathbf{W})$ and leveraging some basic algebra transformations, the beampattern MSE function $\mathcal{E}(\alpha,\mathbf{W})$ is reduced to a univariate function $\mathcal{E}(\mathbf{W})$ and (\ref{eq:optimize_W}b) can be re-arranged as
\be
\label{eq:beampattern1}
\mathcal{E}(\mathbf{W})=\mathrm{vec}^H(\mathbf{W}\mathbf{W}^H)
\mathbf{C}\mathrm{vec}(\mathbf{W}\mathbf{W}^H) \leq \epsilon,
\ee
where we define
\begin{subequations}
\label{eq:C_and_bl}
\begin{align}
\mathbf{C}&\triangleq\frac{1}{L}\sum_{l=1}^{L}\mathbf{b}_l\mathbf{b}_l^H,\\
\mathbf{b}_l&\triangleq\frac{P_\mathrm{d}(\theta_l)
\sum_{l_1=1}^{L}P_\mathrm{d}(\theta_{l_1})\mathrm{vec}(\mathbf{A}_{l_1})}
{\sum_{l_1=1}^{L}P_\mathrm{d}^2 (\theta_{l_1})}-\mathrm{vec}(\mathbf{A}_l).
\end{align}
\end{subequations}

\newcounter{TempEqCnt}
\setcounter{TempEqCnt}{\value{equation}}
\setcounter{equation}{21}
\begin{figure*}[!t]
\begin{subequations}
\begin{align}
\mathbf{B}_1^t&\triangleq\frac{2}{L}
\sum_{l_1=1}^{L}\frac{P_\mathrm{d}^2(\theta_{l_1})}{\beta^2}
\sum_{l_2=1}^{L}P_\mathrm{d}(\theta_{l_2})\mathrm{vec}^H(\mathbf{A}_{l_2})
\mathrm{vec}(\mathbf{W}\mathbf{W}^H)\sum_{l_3=1}^{L}P_\mathrm{d}(\theta_{l_3})\mathbf{A}_{l_3}
+\frac{2}{L}
\sum_{l_1=1}^{L}\mathrm{vec}^H(\mathbf{A}_{l_1})
\mathrm{vec}(\mathbf{W}\mathbf{W}^H)\mathbf{A}_{l_1},\\
\mathbf{B}_2^t&\triangleq-\frac{4}{L}\Re\big\{\!
\sum_{l_1=1}^{L}\frac{P_\mathrm{d}(\theta_{l_1})}{\beta}
\mathrm{vec}^H(\mathbf{A}_{l_1})\mathrm{vec}(\mathbf{W}\mathbf{W}^H)
\sum_{l_2=1}^{L}P_\mathrm{d}(\theta_{l_2})\mathbf{A}_{l_2}\big\}-2\lambda_\mathrm{m}\mathbf{W}^t({\mathbf{W}^t})^H.
\end{align}
\end{subequations}
\rule[-0pt]{18.5 cm}{0.05em}
\end{figure*}

\setcounter{equation}{\value{TempEqCnt}}

Considering the difficulties of tackling the quartic and non-convex beampattern MSE constraint (\ref{eq:beampattern1}), we propose to construct a series of easy-to-optimize surrogate functions for $\mathcal{E}(\mathbf{W})$ by employing the MM method.
Specifically, with the obtained solution $\mathbf{W}^t$ in the $t$-th iteration, a more tractable surrogate function, which approximates $\mathcal{E}(\mathbf{W})$ at the current local point $\mathbf{W}^t$ and serves as an upper bound, is constructed as
\begin{align}
\begin{split}
\label{eq:second^Taylor}
\mathcal{E}(\mathbf{W})& \leq \lambda_\mathrm{m}\mathrm{vec}^H
(\mathbf{W}\mathbf{W}^H)\mathrm{vec}(\mathbf{W}\mathbf{W}^H) \\
&\quad + \Re\{\mathrm{vec}^H(\mathbf{W}\mathbf{W}^H)\mathbf{b}^t\} + c_1^t,
\end{split}
\end{align}
where for brevity we define
\begin{subequations}
\begin{align}
\label{eq:bt}
\mathbf{b}^t&\triangleq2(\mathbf{C}-\lambda_\mathrm{m}\mathbf{I})
\mathrm{vec}(\mathbf{W}^t({\mathbf{W}^t})^H),\\
c_1^t&\triangleq\mathrm{vec}^H(\mathbf{W}^t({\mathbf{W}^t})^H)
(\lambda_\mathrm{m}\mathbf{I}-\mathbf{C})
\mathrm{vec}(\mathbf{W}^t({\mathbf{W}^t})^H),
\end{align}
\end{subequations}
and $\lambda_\mathrm{m}$ denotes the maximum eigenvalue of $\mathbf{C}$.
Thanks to the power constraint (\ref{eq:optimize_W}c), an upper bound of the first quartic term on the right-hand side in (\ref{eq:second^Taylor}) can be easily obtained as
\begin{align}
\begin{split}
\label{eq:second_Taylor1}
&\lambda_\mathrm{m}\mathrm{vec}^H(\mathbf{W}\mathbf{W}^H)
\mathrm{vec}(\mathbf{W}\mathbf{W}^H)=\lambda_\mathrm{m} \Big\| \sum_{j=1}^{K+M}\mathbf{w}_j\mathbf{w}_j^H \Big\|_F^4\\
&\quad\leq\lambda_\mathrm{m}\Big( \sum_{j=1}^{K+M}\| \mathbf{w}_j\mathbf{w}_j^H \|_F^2 \Big)^2
=\lambda_\mathrm{m}P^2.
\end{split}
\end{align}
Substituting (\ref{eq:second_Taylor1}) into (\ref{eq:second^Taylor}), an upper bound of the beampattern MSE function $\mathcal{E}(\mathbf{W})$ can be thus expressed as
\begin{equation}
\label{eq:upper_bound1}
\mathcal{E}(\mathbf{W})\leq\Re\{ \mathrm{vec}^H(\mathbf{W}\mathbf{W}^H)\mathbf{b}^t \} + c_1^t + \lambda_\mathrm{m} P^2.
\end{equation}

To facilitate the algorithm development, we equivalently and explicitly re-write the term $\Re\{\mathrm{vec}^H(\mathbf{W}\mathbf{W}^H)\mathbf{b}^t\}$ as
\begin{equation}
\begin{split}
\label{eq:bt2Bt}
\Re\{\mathrm{vec}^H(\mathbf{W}\mathbf{W}^H)\mathbf{b}^t\}
=&\sum_{j=1}^{K+M}\Re\{\mathrm{vec}^H(\mathbf{w}_j\mathbf{w}_j^H)\mathbf{b}^t\}\\
=&\sum_{j=1}^{K+M}\Re\{\mathbf{w}_j^H\mathbf{B}^t\mathbf{w}_j\},
\end{split}
\end{equation}
where $\mathbf{B}^t \in \mathbb{C}^{N \times N}$ is a reshaped version of $\mathbf{b}^t$, i.e., $\mathbf{b}^t=\mathrm{vec}(\mathbf{B}^t)$.
Based on the definitions in (\ref{eq:C_and_bl}) and (\ref{eq:bt}), the matrix $\mathbf{B}^t$ can be split into two parts $\mathbf{B}^t\triangleq\mathbf{B}_1^t+\mathbf{B}_2^t$ with $\mathbf{B}_1^t$ and $\mathbf{B}_2^t$ being defined on the top of the next page, where for brevity we define the scalar $\beta\triangleq\!\sum_{l=1}^{L}P_\mathrm{d}^2(\theta_{l})$.
It is obvious that the matrix $\mathbf{B}^t$ equals to a positive semidefinite matrix $\mathbf{B}_1^t$ plus a negative semidefinite matrix $\mathbf{B}_2^t$.
In other words, the function $\mathbf{w}_j^H\mathbf{B}^t\mathbf{w}_j$ in (\ref{eq:bt2Bt}) can be separated into a convex function $\mathbf{w}_j^H\mathbf{B}_1^t\mathbf{w}_j$ plus a concave function $\mathbf{w}_j^H\mathbf{B}_2^t\mathbf{w}_j$.
In order to handle the non-convex part $\mathbf{w}_j^H\mathbf{B}_2^t\mathbf{w}_j$, we employ the MM method to find a convex surrogate function for it in each iteration.
In particular, by exploiting the first-order Taylor expansion, a convex upper bound of the concave function $\mathbf{w}_j^H\mathbf{B}_2^t\mathbf{w}_j$ can be constructed as
\be
\setcounter{equation}{23}
\label{eq:B2t}
\mathbf{w}_j^H\mathbf{B}_2^t\mathbf{w}_j\leq ({\mathbf{w}_j^t})^H\mathbf{B}_2^t\mathbf{w}_j^t+2\Re\big\{({\mathbf{w}_j^t})^H
\mathbf{B}_2^t(\mathbf{w}_j-\mathbf{w}_j^t)\big\},
\ee
where $\mathbf{w}_j^t$ is the $j$-th column of $\mathbf{W}^t$ obtained in the $t$-th iteration.
Therefore, plugging the results in (\ref{eq:bt2Bt})-(23) into (\ref{eq:upper_bound1}), a convex upper bound of the beampattern MSE function  $\mathcal{E}(\mathbf{W})$ can be obtained as
\be
\label{eq:first^Taylor}
\mathcal{E}(\mathbf{W})\leq\sum_{j=1}^{K+M}
\Re\big\{\mathbf{w}_j^H\mathbf{B}_1^t\mathbf{w}_j
+2\mathbf{w}_j^H\mathbf{u}_j\big\}+c_2^t,
\ee
where for brevity we define
\begin{subequations}
\begin{align}
\mathbf{u}_j&\triangleq ({\mathbf{B}_2^t})^H\mathbf{w}_j^t,\\
c_2^t&\triangleq-\!\!\!\sum_{j=1}^{K+M}
\Re\{({\mathbf{w}_j^t})^H({\mathbf{B}_2^t})^H\mathbf{w}_j^t\}+c_1^t + \lambda_\text{m}P^2.
\end{align}
\end{subequations}

With the surrogate function of the beampattern MSE function $\mathcal{E}(\mathbf{W})$ in (\ref{eq:first^Taylor}), the optimization problem of updating $\mathbf{W}$ can be reformulated as
\begin{subequations}
\label{eq:opt_W}
\begin{align}
\max_{\mathbf{W}}&\,\sum\limits_{k=1}^{K}\!\big(2\sqrt{1+c_k}\Re\{ g_k^{*}\mathbf{h}_k^H \mathbf{w}_k\!\}\!-\!| g_k |^2\!\!\! \sum\limits_{j=1}^{K+M}\!\!\!| \mathbf{h}_k^H \mathbf{w}_j |^2\big)\\
\text{s.t.}~&\!\sum_{j=1}^{K+M}\Re\{ \mathbf{w}_j^H\mathbf{B}_1^t\mathbf{w}_j+2\mathbf{w}_j^H\mathbf{u}_j\}+c_2^t
\leq\epsilon,\\
&\;\| \mathbf{W} \|_F^2 \leq P.
\end{align}
\end{subequations}
We clearly find that problem (\ref{eq:opt_W}) is convex and can be easily solved by existing convex optimization solvers such as CVX.

\subsection{Optimize $\pmb{\phi}$ with Given $\mathbf{W}$}

After obtaining the beamforming matrix $\mathbf{W}$, the optimization problem of solving $\pmb{\phi}$ can be formulated as
\begin{subequations}
\label{eq:optimize_Phi}
\begin{align}
\max_{\pmb{\phi}}&\,\sum\limits_{k=1}^{K}\!\big(2\sqrt{1+c_k}\Re\{ g_k^{*}\mathbf{h}_k^H \mathbf{w}_k\!\}\!-\!| g_k |^2\!\!\! \sum\limits_{j=1}^{K+M}\!\!\!| \mathbf{h}_k^H \mathbf{w}_j |^2\big)\\
\text{s.t.}~&\,\left| \phi_n \right|=1,~\forall n.
\end{align}
\end{subequations}
Then, utilizing the equation $\mathbf{h}_{\mathrm{r},k}^H\pmb{\Phi}\mathbf{G}\mathbf{w}_j=
\mathbf{h}_{\mathrm{r},k}^H\mathrm{diag}(\mathbf{G}\mathbf{w}_j)\pmb{\phi}$, problem (\ref{eq:optimize_Phi}) can be re-arranged as
\begin{subequations}
\label{eq:opt_phase}
\begin{align}
\min_{\pmb{\phi}}~~&\pmb{\phi}^H\mathbf{Q}\pmb{\phi}-2\Re\{\pmb{\phi}^H\mathbf{q}\}-c\\
\text{s.t.}~~&\left| \phi_n \right|=1,~\forall n,
\end{align}
\end{subequations}
where for brevity we define
\begin{subequations}\begin{align}
\mathbf{Q}&\triangleq\!\sum\limits_{k=1}^{K}\!|g_k|^2\!\! \sum\limits_{j=1}^{K+M}\!\!\mathrm{diag}(\!\mathbf{w}_j^H\mathbf{G}^H\!)
\mathbf{h}_{\mathrm{r},k}\mathbf{h}_{\mathrm{r},k}^H
\mathrm{diag}(\!\mathbf{G}\mathbf{w}_j\!),\!\!\\
\mathbf{q}&\triangleq\!\sum\limits_{k=1}^{K}\sqrt{1+c_k}g_k
\mathrm{diag}(\mathbf{w}_k^H\mathbf{G}^H)\mathbf{h}_{\mathrm{r},k}
\non\\
&~~~\;-\sum\limits_{k=1}^{K}|g_k|^2\!\!\sum\limits_{j=1}^{K+M}\mathrm{diag}(\mathbf{w}_j^H\mathbf{G}^H)\mathbf{h}_{\mathrm{r},k}
\mathbf{h}_{\mathrm{d},k}^H\mathbf{w}_j^H,\\
c&\triangleq\Re\Big\{\!\sum\limits_{k=1}^{K}\!2\sqrt{1+c_k}g_k^{*}
\mathbf{h}_{\mathrm{d},k}^H\mathbf{w}_k\!-\!|g_k|^2\!\!\!\sum\limits_{j=1}^{K+M}
\!\!\!|\mathbf{h}_{\mathrm{d},k}^H\mathbf{w}_j|^2\!\Big\}.\!\!
\end{align}
\end{subequations}

We observe that the major difficulty to cope with problem (\ref{eq:opt_phase}) is the non-convex unit-modulus constraint (\ref{eq:opt_phase}b).
Considering the performance and effectiveness of various existing algorithms in solving problem (\ref{eq:opt_phase}), we employ the manifold-based algorithm in this paper.
Specifically, the objective function (\ref{eq:opt_phase}a) is smooth and the unit-modulus constraint (\ref{eq:opt_phase}b) forms a complex circle Riemannian manifold, which allows problem (\ref{eq:opt_phase}) to be solved with the typical Riemannian conjugate gradient (RCG) algorithm.
After deriving the Riemannian gradient from the corresponding Euclidean gradient, problem (\ref{eq:opt_phase}) can be iteratively solved on the Riemannian space by utilizing the idea of conjugate gradient algorithm.
Readers can refer to \cite{Rang TWC 2021} for more details.

\section{Simulation Results}
\begin{figure*}[htbp]
\centering
\begin{minipage}[t]{0.33\linewidth}
\setcaptionwidth{1.8 in}
\centering
\includegraphics[width=2.53 in,height=2 in]{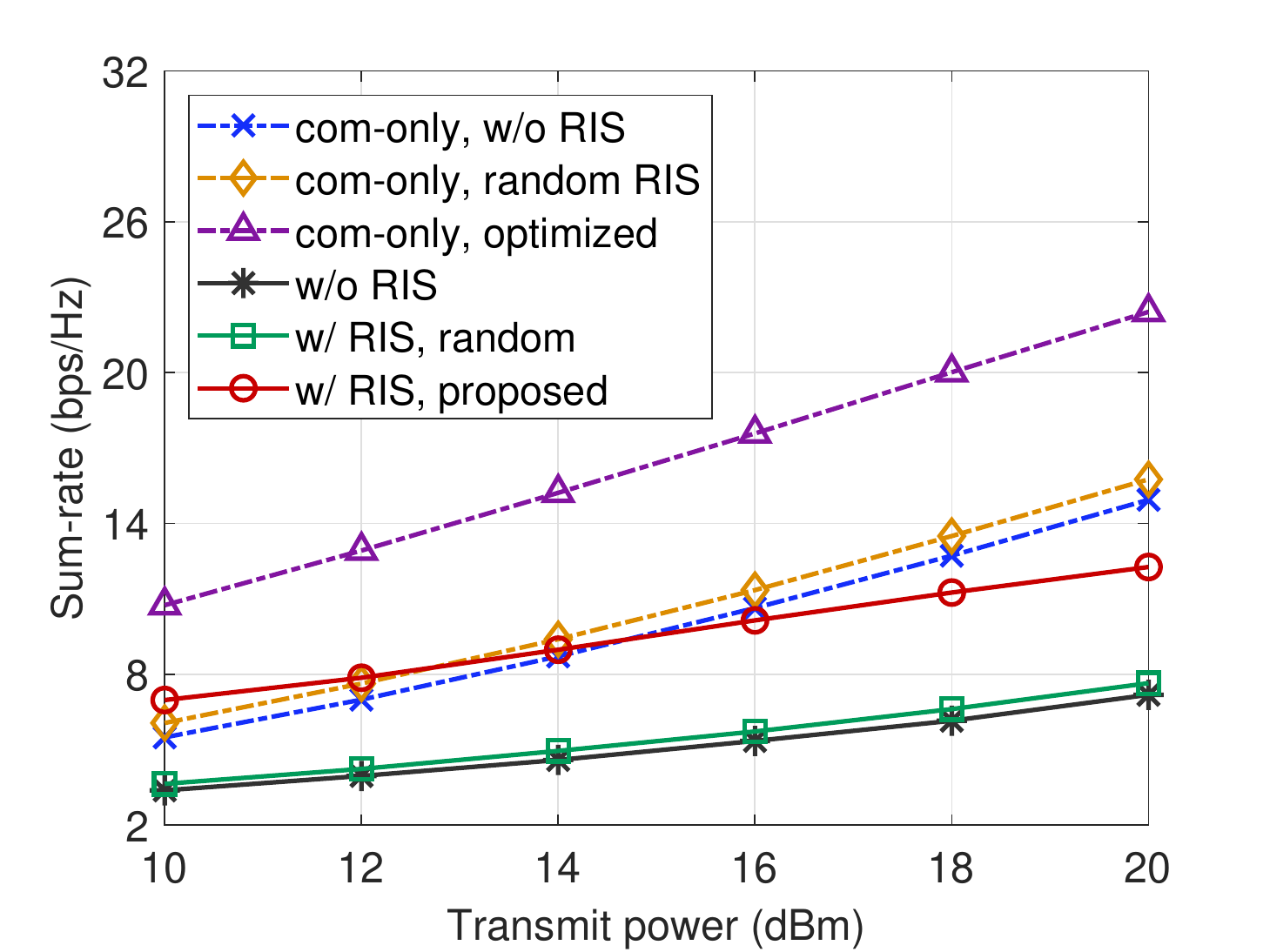}
\caption{Sum-rate versus transmit power ($N=169$).}
\label{fig:power}
\end{minipage}%
\begin{minipage}[t]{0.33\linewidth}
\setcaptionwidth{2.1 in}
\centering
\includegraphics[width=2.53 in,height=2 in]{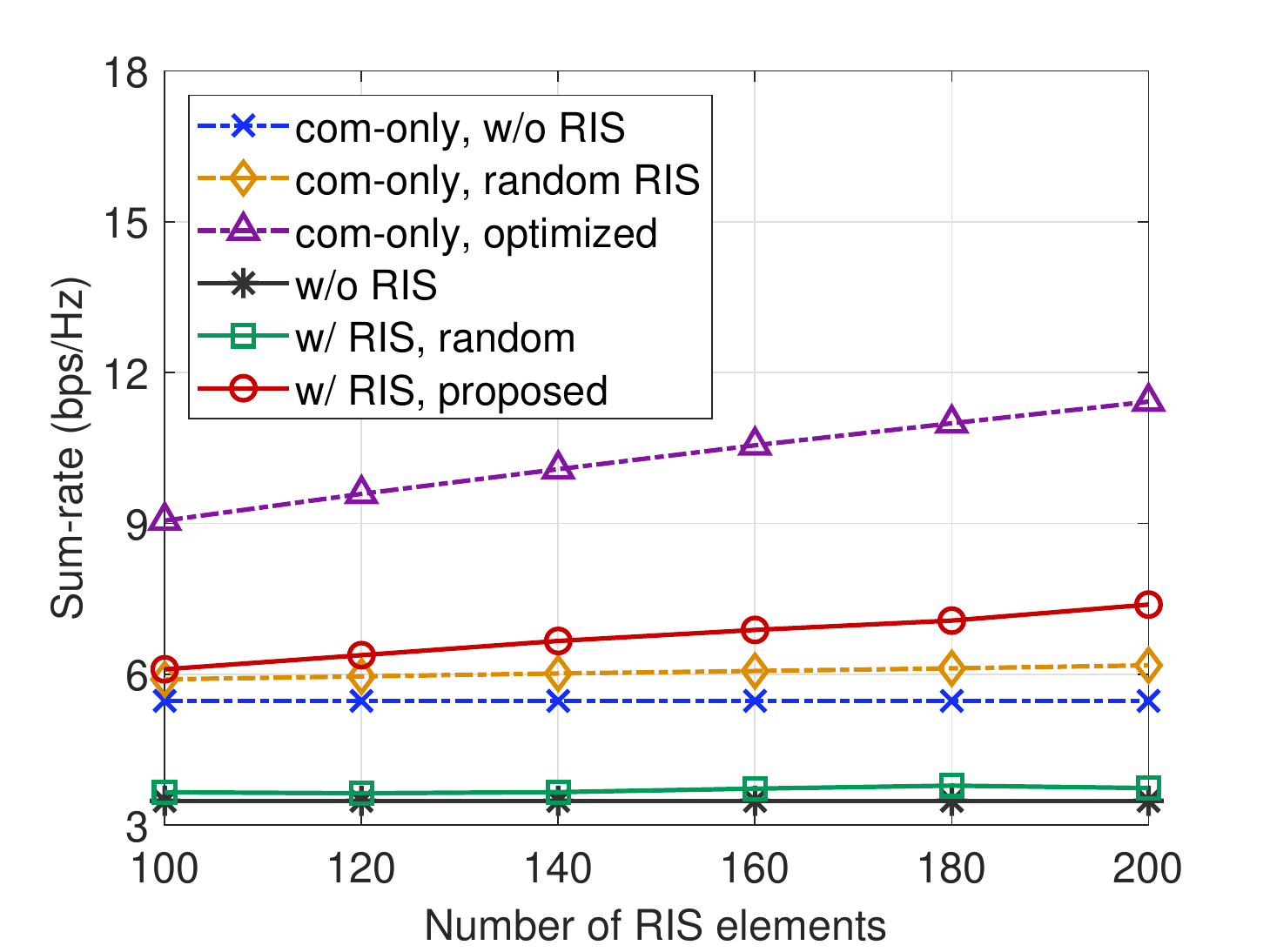}
\caption{Sum-rate versus the number of RIS elements ($P=10\mathrm{dBm}$).}
\label{fig:RIS_elements}
\end{minipage}%
\begin{minipage}[t]{0.33\linewidth}
\centering
\includegraphics[width=2.53 in,height=2 in]{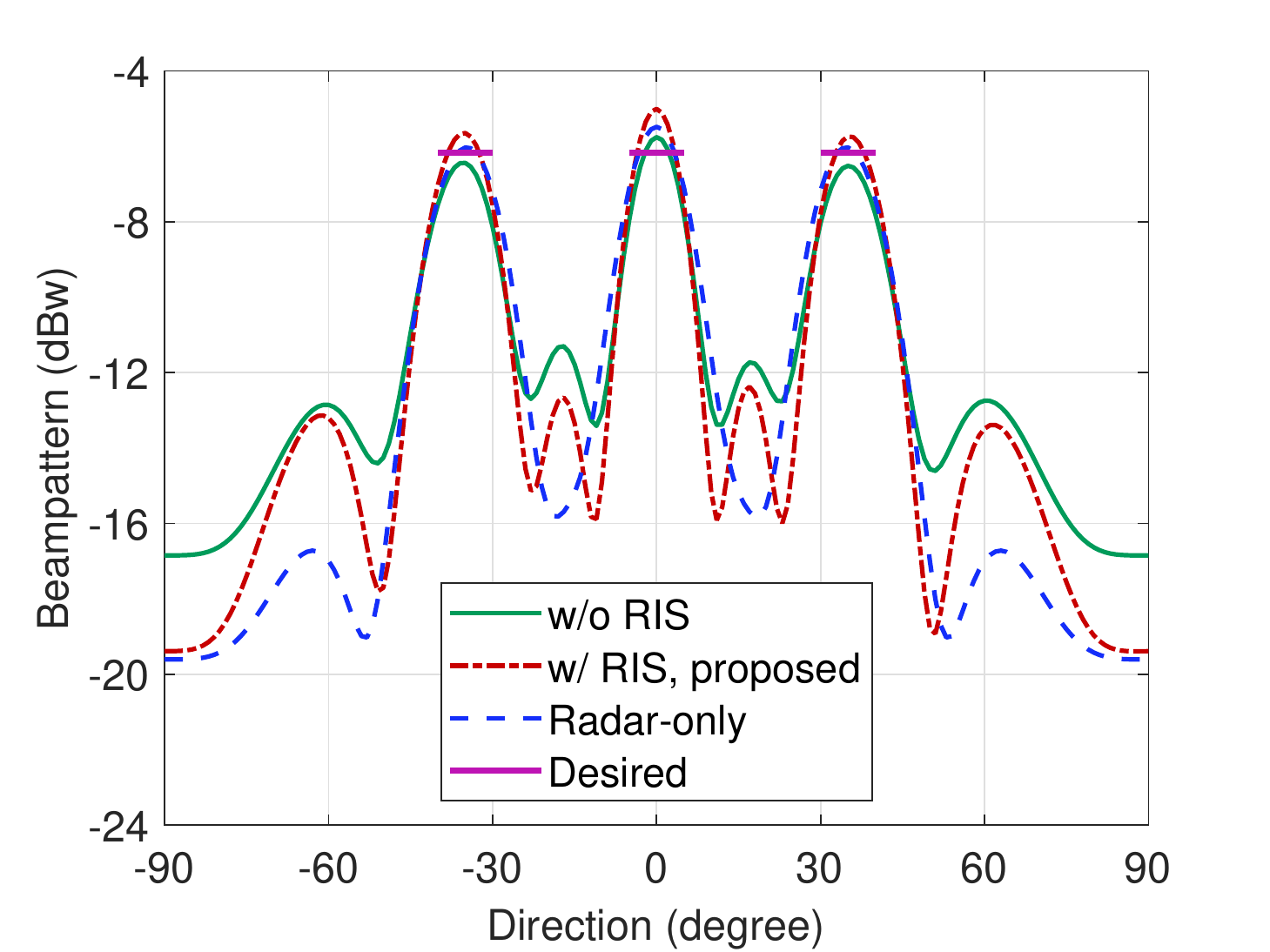}
\caption{Transmit beampattern.}
\label{fig:beampattern}
\end{minipage}
\centering
\end{figure*}

In this section, we numerically evaluate the performance of our proposed joint beamforming design for the RIS-assisted ISAC system.
We assume that the BS equipped with $M=8$ antennas in half-wavelength antenna spacing serves $K=4$ single-antenna users with the assistance of an $N$-element RIS.
The noise power of users is set as $\sigma_k^2=-80\mathrm{dBm},~\forall k$.
The distances of BS-RIS and RIS-user links are set as 50m and 4m, respectively.
We adopt the typical distance-dependent path-loss model \cite{Wu TWC 2019} and set the path-loss exponents of the BS-user, BS-RIS, and RIS-user links as $\alpha_\mathrm{Bu}=3.5$, $\alpha_\mathrm{BR}=2.5$, and $\alpha_\mathrm{Ru}=2.5$, respectively.
The Rician fading channel model is assumed with the Rician factors being $\beta_\mathrm{BR}=\beta_\mathrm{Ru}=3\mathrm{dB}$ and $\beta_\mathrm{Bu}=0$.
Besides, the ideal beampattern $P_\mathrm{d}(\theta_l)$ is given by
\begin{equation}
P_\mathrm{d}(\theta_l)=
\begin{cases}
1,\theta_l\in[\overline{\theta}_t-\frac{\Delta_\theta}{2},\overline{\theta}_t+\frac{\Delta_\theta}{2}],t=1,\cdots\!,T,\\
0,\mathrm{otherwise},
\end{cases}
\end{equation}
where $\overline{\theta}_t$ represents the direction of the $t$-th target, $\overline{\theta}_1=-35^\circ, \overline{\theta}_2=0^\circ, \overline{\theta}_3=35^\circ$, and $\Delta_\theta=10^\circ$ is the beam width.

We first illustrate the achievable sum-rate versus the transmit power in Fig. \ref{fig:power}, where the scenarios without RIS (``w/o RIS'') and with random phase-shift RIS (``w/ RIS, random'') are included in addition to the proposed algorithm (``w/ RIS, proposed'').
Besides, a communication-only (``com-only'') system is also included for comparison.
It is easy to find that the scenarios with RIS achieve better performance than those without RIS, since the RIS introduces additional NLoS links that can enhance the downlink communication.
Moreover, the performance improvement provided by the ``w/ RIS, proposed'' scheme is much larger than that of the ``w/ RIS, random'' scheme, which illustrates the effectiveness of the proposed algorithm for jointly designing the transmit beamforming and reflection coefficients.
Additionally, the performance gap between the considered ISAC system and the communication-only system can be observed due to the trade-off between the communication and radar sensing performance.

Then, we present the achievable sum-rate versus the number of RIS elements in Fig. \ref{fig:RIS_elements}.
We clearly observe that a RIS with more reflecting elements offers larger sum-rate, considering that it can exploit more spatial DoFs to achieve higher beamforming gains.
In addition, with the number of reflecting elements increasing, the performance gap between the ``w/ RIS, proposed'' scheme and the ``w/ RIS, random'' scheme becomes larger, which implies the significance of optimizing a large-scale RIS in pursuing performance improvement.

Finally, we plot the transmit beampatterns to show the radar sensing performance in Fig. \ref{fig:beampattern}, where the desired ideal beampattern (``Desired'') and the achieved beampattern in practical radar-only scenario \cite{J. Li} (``Radar-only'') are also included.
Compared with the ``Radar-only'' benchmark, the transmit beampattern of the proposed scheme has higher sidelobes due to the additional communication functionality. However, it has a more preferable shape than that of the ``w/o RIS'' scheme, which verifies that the ISAC system possesses better sensing performance by deploying RIS to assist communications.

\section{Conclusion}
In this paper, we investigated the deployment of RIS in ISAC systems.
The transmit beamforming and the reflection coefficients were jointly optimized to maximize the achievable sum-rate as well as guarantee the constraints of the beampattern similarity, the total transmit power, and the phase-shift of the RIS.
An efficient alternating algorithm was developed to solve the resulting problem.
Simulation results demonstrated the significant advantages of deploying RIS in ISAC systems.
We will extend our work to more general scenarios where there is clutter or blocked target, as well as other crucial issues including the performance trade-off and fairness.

\end{document}